# Science Potential of a Deep Ocean Antineutrino Observatory

S.T. Dye

Department of Physics and Astronomy, University of Hawaii
2505 Correa Road, Honolulu, Hawaii, 96822 U.S.A.

College of Natural Sciences, Hawaii Pacific University
45-045 Kamehameha Highway, Kaneohe, Hawaii 96744 U.S.A.

This paper presents science potential of a deep ocean antineutrino observatory being developed at Hawaii. The observatory design allows for relocation from one site to another. Positioning the observatory some 60 km distant from a nuclear reactor complex enables precision measurement of neutrino mixing parameters, leading to a determination of neutrino mass hierarchy and $\theta_{13}$. At a mid-Pacific location the observatory measures the flux and ratio of uranium and thorium decay neutrinos from earth's mantle and performs a sensitive search for a hypothetical natural fission reactor in earth's core. A subsequent deployment at another mid-ocean location would test lateral heterogeneity of uranium and thorium in earth's mantle.

## 1. INTRODUCTION

A deep ocean antineutrino observatory called Hanohano (Hawaii Anti-Neutrino Observatory) is being developed at Hawaii. The observatory records interactions of electron antineutrinos of $E_\nu$>1.8 MeV by inverse β-decay in a monolithic cylindrical detector of 10 kt of ultra-pure scintillating liquid. An outer surface array of inward-looking 10-inch photomultiplier tubes in 13-inch glass pressure housings collects scintillation photons. An energy resolution of 3.5%/$\sqrt{E_{vis}}$ results, with $E_{vis}$=$E_\nu$-0.8 MeV the visible energy. Sufficient overburden, adequate shielding, and radio-pure detector components limit background to negligible levels providing very high detection efficiency. The observatory does not accurately measure the direction of antineutrinos.

Considerable science potential derives from the ability to deploy the observatory at various deep ocean locations. An initial deployment offshore a nuclear reactor complex for measuring neutrino mixing parameters could be followed, for example, by a deployment near Hawaii for measuring terrestrial antineutrinos. This flexibility presents a significant advantage over a similar observatory at a fixed underground location.

## 2. NEUTRINO MIXING PARAMETERS

Neutrino mixing and oscillation [1] are responsible for the deficit of solar neutrinos [2], the spectral distortion of reactor antineutrinos [3], and the deficit of atmospheric muon neutrinos [4] confirmed using an accelerator-produced muon neutrino beam [5]. These initial observations reduce the allowed regions of neutrino mixing parameter space guiding future precision measurements of mixing angles and mass-squared differences, including resolution of the spectrum of neutrino masses. Positioning Hanohano ~60 km distant from a nuclear reactor complex enables precision measurement of $\theta_{12}$ and, for non-zero $\theta_{13}$, $\delta m^2_{31}$. This latter measurement leads to a determination of neutrino mass hierarchy.

Several authors discuss a precision measurement of the solar mixing angle $\theta_{12}$ using antineutrinos from a nuclear reactor [6-8]. The experiment utilizes a near detector at the reactor complex for normalizing flux and cross section

hep-ex/0611039

with a far detector at the first minimum of survival probability. There is agreement that an exposure of 60 GW-kt-y of a far detector at a distance of ~60 km yields an uncertainty of 2% in the value of $\sin^2(\theta_{12})$ at the 68% confidence level, assuming a detector systematic uncertainty of 4% or less. This experiment is analogous to methods proposed for precision measurement of the sub-dominant mixing angle $\theta_{13}$ [9], namely the reactor antineutrino flux sampling defines one-half cycle of the oscillating survival probability. The difference is the distance of the first minimum of survival probability, which is ~2 km for the $\theta_{13}$ measurement.

The far detector for the $\theta_{12}$ experiment records multiple cycles of $\delta m^2_{31}$ oscillation provided $\theta_{13} \neq 0$, adequate energy resolution, and sufficient exposure. There is a plan to measure these cycles in $L/E$ space by sampling the Fourier power at different values of $\delta m^2$ [10]. This self-normalizing, robust method offers a precision measurement of $\delta m^2_{31}$ for $\sin^2(2)>0.05$, determines neutrino mass hierarchy by evaluating asymmetry of the Fourier power spectrum, and measures $\theta_{13}$ all without the need for a near detector.

Hanohano is capable of performing the experiments described above with a one-year deployment offshore a suitable nuclear reactor complex. At least two candidate sites with considerable overburden exist. One is in 1100 m of water west of the ~7 GW San Onofre reactors in California and the other is in 2800 m of water east of the ~6 GW Maanshan reactors in Taiwan.

## 3. TERRESTRIAL ANTINEUTRINOS

Numerous authors have discussed measuring antineutrinos from β-decays of long-lived radioactive isotopes in the earth to study earth composition and energetics [11-16]. Recently an experiment has detected the energy spectrum but not the direction of geo-neutrinos from uranium and thorium decays [17]. Considerable scientific interest has ensued, resulting in an observational program of refined measurements and new initiatives [18].

Hanohano is a new initiative capable of measuring the flux of U/Th geo-neutrinos from earth's mantle with 25% uncertainty in one year of operation near Hawaii [19]. Included in this statistics dominated result is 9% systematic error due to uncertainty in the U/Th content of the crusts. This same uncertainty limits the precision of measurements of the mantle flux at continental locations to >50%. Not included in the analysis is uncertainty of the neutrino mixing angles $\theta_{12}$ [3] and $\theta_{13}$ [20]. The survival probability for fully mixed geo-neutrinos is

$$P_{(\bar{\nu}_e \to \bar{\nu}_e)} = 1 - \frac{1}{2}\left[\cos^4(\theta_{13})\sin^2(2\theta_{12}) + \sin^2(2\theta_{13})\right]$$
$$= 0.59 \cdot (1.00^{+0.06}_{-0.15})$$

The upper (lower) value obtains with minimum (maximum) values of the mixing angles. Imprecise knowledge of mixing angles and U/Th content of earth's crusts introduce comparable uncertainties to the measurement by Hanohano of geo-neutrinos from the mantle. Nonetheless, deployments at several widely-spaced mid-ocean locations test lateral heterogeneity of uranium and thorium in the mantle.

Geo-neutrinos with energy between 1.8 MeV and 2.3 MeV come from both $^{238}$U and $^{232}$Th decay products, while those between 2.3 MeV and the maximum energy of 3.3 MeV are only from the $^{238}$U decay product $^{214}$Bi. This spectral feature allows a measurement of the Th/U ratio. Although geology traditionally ascribes the chondritic Th/U ratio of 3.9±0.1 to the bulk earth, samples from the upper mantle reveal a substantially lower value of 2.6 suggesting layered mantle convection [21]. Geo-neutrino flux measurements sample large volumes of the deep earth. A four-year exposure of Hanohano near Hawaii measures the Th/U ratio of the sampled earth to 10%, providing an important test of mantle convection models.

An earth-centered natural fission reactor [22,23] is a speculative, untested hypothesis. Predicted to be in the power range of 1-10 TW, it has the potential to explain the variability of the geo-magnetic field and the anomalously high helium-3 concentrations in hot-spot lavas. Fission products from such a geo-reactor would undergo β-decay, producing antineutrinos with the characteristic nuclear reactor spectrum. A one-year deployment of Hanohano at a mid-ocean location well distant from nuclear power plants tests the existence of the geo-reactor. This deployment would set a 99% CL upper limit to the geo-reactor power at 0.3 TW or, were a 1 TW geo-reactor to exist, produce nearly a 5σ measurement [19].

4. SUMMARY

Hanohano is a deep ocean antineutrino observatory being developed at Hawaii. A one-year deployment ~60 km distant from a nuclear reactor complex measures $\sin^2(\theta_{12})$ to 2%. This measurement requires a dedicated near detector. The same deployment measures $\delta m^2_{31}$ and determines neutrino mass hierarchy for $\sin^2(2\theta_{13})>0.05$ without a near detector. A one-year deployment near Hawaii measures the flux of U/Th geo-neutrinos from the mantle to 25% and either measures or severely constrains the power of the hypothetical geo-reactor. An exposure of four years measures the Th/U ratio to 10%.

This work was partially funded by Hawaii Pacific University and U.S. Department of Energy grant DE-FG02-04ER41291.